\documentclass[a4paper,nofootinbib,preprintnumbers,twocolumn,preprintnumbers,floatfix,superscriptaddress,pra,showpacs]{revtex4-1}

\usepackage{graphicx}
\usepackage{amssymb}
\usepackage{amsmath}
\usepackage{epstopdf}
\usepackage{graphics}
\usepackage[caption=false]{subfig}
\usepackage{float}
\usepackage{color}
\usepackage{hyperref}

\newcommand{\ket}[1]{|#1\rangle}					
\newcommand{\braket}[2]{\langle #1 | #2 \rangle}	

\newcommand{\matrixel}[3]{\left< #1 \vphantom{#2#3} \right| #2 \left| #3 \vphantom{#1#2} \right>} 
\newcommand{\abs}[1]{\left| #1 \right|} 

\newcommand{\eqnref}[1]{(\ref{#1})}
\newcommand{\figref}[1]{Fig.~\ref{#1}}
\newcommand{\tabref}[1]{Table \ref{#1}}
\newcommand{\secref}[1]{Sec.~\ref{#1}}
\newcommand{\appref}[1]{App.~\ref{#1}}

\graphicspath{{../Figs/}}

\begin{document}

\title{Hybrid Quantum Repeater Protocol With Fast Local Processing}

\author{J. Borregaard }
\affiliation{QUANTOP, Niels Bohr Institute, Blegdamsvej 17, DK-2100 Copenhagen ¯, Denmark}
\author{J. B. Brask}
\affiliation{ICFO-Institut de Ciencies Fotoniques, Av.~Carl Friedrich Gauss, 3, 08860 Castelldefels (Barcelona), Spain}
\author{A. S. S\o rensen}
\affiliation{QUANTOP, Niels Bohr Institute, Blegdamsvej 17, DK-2100 Copenhagen ¯, Denmark}

\date{\today}

\begin{abstract}
We propose a hybrid quantum repeater protocol combining the advantages of continuous and discrete variables. The repeater is based on the previous work of Brask \textit{et al.} [Phys. Rev. Lett. 105, 160501 (2010)] but we present two ways of improving this protocol. In the previous protocol entangled single-photon states are produced and grown into superpositions of coherent states, known as two-mode cat states. The entanglement is then distributed using homodyne detection. To improve the protocol, we replace the time-consuming non-local growth of cat states with local growth of single-mode cat states, eliminating the need for classical communication during growth. Entanglement is generated in subsequent connection processes. Furthermore the growth procedure is optimized. We review the main elements of the original protocol and present the two modifications. Finally the two protocols are compared and the modified protocol is shown to perform significantly better than the original protocol. 
\end{abstract}


\maketitle

\section{Introduction}
\label{sec.intro}

A major goal in the field of quantum information is distributing entanglement over large distances. A strong motivation for this is that it may enable transmission of information, which is secure against eavesdropping~\cite{ekert1991,gisin2002} even in cases where the measurement devices or the source are untrusted \cite{acin2007}. More generally the distribution of entanglement is required for almost any task in quantum communication.  Direct distribution of entanglement requires transmission of fragile quantum states, which is difficult in practice due to loss and decoherence in optical fibers. Quantum repeaters overcome this problem by initially generating entanglement over short distances and then distributing it via entanglement swapping, which only requires local operations and classical communication~\cite{wootters1982}. Recently, much effort has been devoted to the construction of quantum repeaters based on atomic ensembles \cite{duan2001, zhao2008,zhao2007,yuan2008,chen2008,kuzmich2006,laurat2007,choi2008,chou2005,kimble2008,sangouard2008,simon2007,sangouard2010,sangouard2011,usmani2010}. Despite the experimental advances towards this goal the construction of  a fully functioning quantum  repeater remain a formidable challenge. In particular the low efficiencies obtained in current experiments severely reduce the achievable communication rates. Numerous theoretical proposals have been developed to improve the communication rates \cite{sangouard2011,sangouard2009,jiang2007,zhao2007,chen2007}. In particular people have suggested schemes relying on multiplexing of operations so that high local repetition rates can compensate for the slow non-local operation which require communication with distant parties \cite{simon2007,collins2007,simon2010pra}. 

Quantum communication generally works in two regimes; the discrete and the continuous variable regime. In the discrete variable regime, information is carried by single photons and measurements rely on single photon detection (SPD). This facilitates detection and correction for loss, but the efficiency of most available single-photon detectors is low, reducing the rate of entanglement distribution. High-efficiency ($>90\%$) SPD is possible, but requires detectors, such as superconducting transition-edge sensors, which are expensive and not widely available \cite{hadfield2009}. In the continuous variable regime, information is encoded in operators with a continuous spectrum such as the field quadratures of the electromagnetic field. These are measured using homodyne detection which is very efficient in practice ($\sim99\%$) but has the drawback that loss is not as easily detected as in the discrete variable regime. Recently hybrid quantum repeater protocols, combining the two regimes, were proposed first for spin systems in cavities \cite{loock2006} and later for atomic ensembles ~\cite{brask2010prl}. Here we follow the approach of Ref.~\cite{brask2010prl} for atomic ensembles. The performance of this repeater protocol is comparable to the best proposed atomic-ensemble based repeaters in the discrete variable regime if these are operated using realistic SPD with limited efficiency~\cite{duan2001,sangouard2011,sangouard2010}.

Here we propose and analyze two modifications to the protocol of Ref.~\cite{brask2010prl}. The hybrid repeater protocol creates entanglement between two stations in the form of single-photon superpositions $\ket{01} + \ket{10}$ ignoring normalization for simplicity. Through a probabilistic procedure, these states are then grown into states resembling
\begin{equation}
\label{eq:cat}
\ket{\gamma(\theta,\alpha)}\varpropto e^{i\theta}\ket{\alpha}_{a}\ket{\alpha}_{b}+e^{-i\theta}\ket{-\alpha}_{a}\ket{-\alpha}_{b},
\end{equation}
by means of local operations and classical communication. Here $\ket{\alpha}_{a}$ denotes a coherent state with amplitude $\alpha$ in mode $a$, and $\theta$ is a phase. We refer to states of the form in Eq. \eqref{eq:cat} as two-mode cat states since they are two mode superpositions of two "classical" states $\ket{\alpha},\ket{-\alpha}$. Because classical communication between distant stations is time consuming, the growth procedure is slow.  In particular the single-photon entanglement generation step, which has low success probability, needs to be repeated every time the growth step fails. In a related setup a solution to this problem was suggested in Ref. \cite{sangouard2008}.  To improve the communication rate, it was proposed to replace the low success non-local entanglement generation by a rapid preparation of a suitable local states.  Because local operations do not rely on communication with distant parties they have a much higher obtainable rate. The locally generated states are more suitable for entanglement generation and can be connected with a much higher probability reducing the time spent on the slow non-local operations.  Here we follow a similar path and consider interchanging the first two steps of the repeater protocol such that states resembling one-mode cat states,
\begin{equation}
\label{eq:singcat}
\ket{\xi(\theta,\alpha)}\varpropto e^{i\theta}\ket{\alpha}+e^{-i\theta}\ket{-\alpha} ,
\end{equation}
are first grown locally and then subsequently connected to create entanglement by means of non-local single-photon subtraction. Such a modification reduces the need for classical communication and allows a higher repetition rate to be reached. This is the main idea behind the new repeater protocol detailed below. In addition, we optimize the cat-state growth procedure, improving the rate further. 
The remainder of the paper is organised as follows. In \secref{sec.prevrep} we review the scheme of Ref.~\cite{brask2010prl}. In \secref{sec.newrep} we present the modified protocol. In \secref{sec.perform} we present results of numerical simulation of the new scheme, evaluating the performance in terms of achievable rates. Finally we conclude in \secref{sec.conclusion}. Some detailed calculations and parameters relevant to the numerics are given in appendices \ref{sec:appgrowing}, \ref{sec:appconnect}, \ref{sec:appswap} and \ref{app:supma}.

\section{Review of previous scheme}
\label{sec.prevrep}

\begin{figure} [t,p]
\centering
\includegraphics[width=0.5\textwidth]{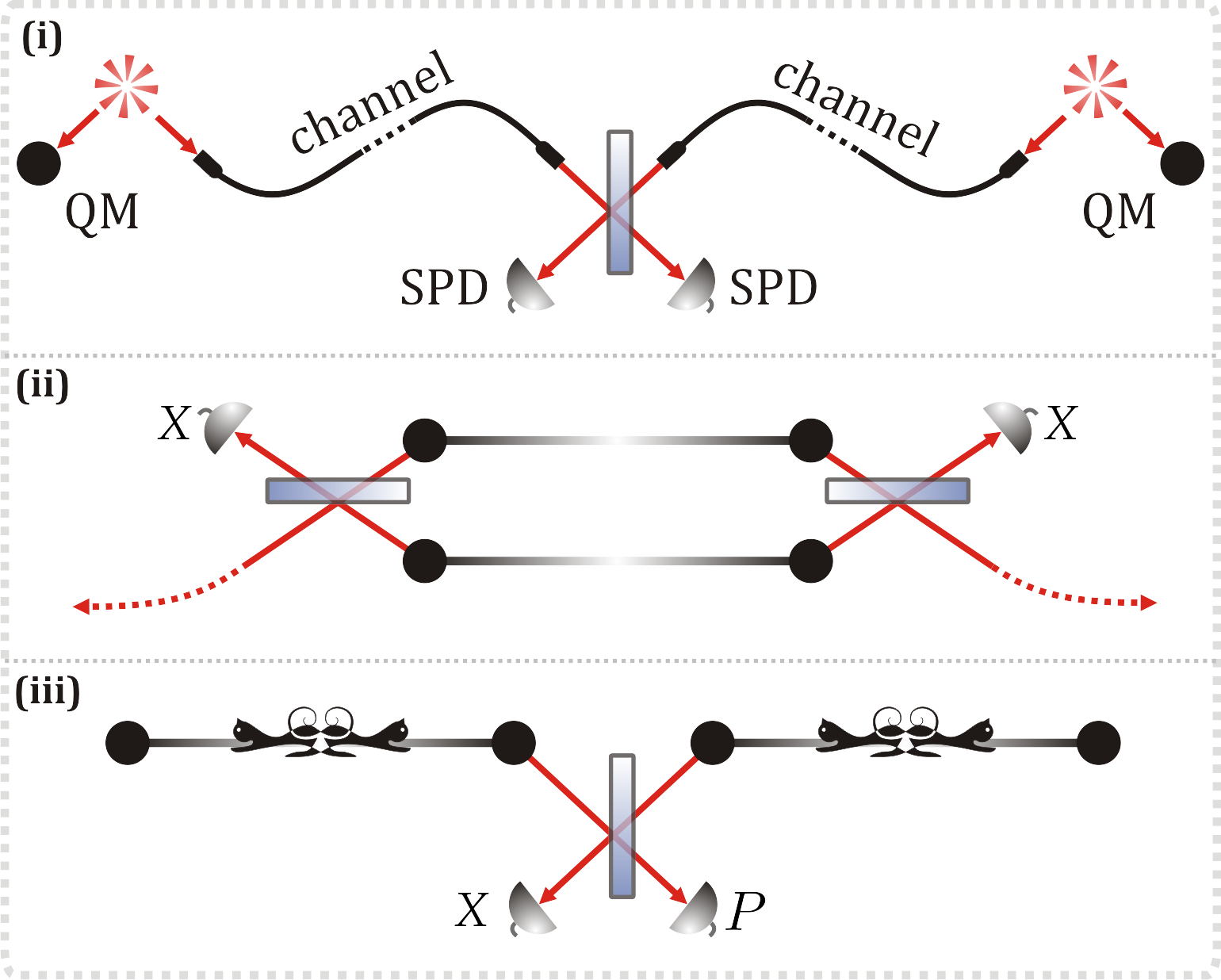}
\caption{(Colour online) Steps of the protocol in Ref.~\cite{brask2010prl}. \textbf{(i)} Entanglement is generated using two sources of two-mode squeezed vacuum. One mode from each source is transmitted to a balanced beam splitter and the outputs are measured. Detection of a single photon heralds entanglement between the remaining modes stored in quantum memories (QM). \textbf{(ii)} Growth of cat states. Two entangled states are combined locally on balanced beam splitters and the $\hat{X}$ quadrature is measured. Success is conditioned on the sum of the outcomes taking a value close to zero. \textbf{(iii)} Entanglement swapping. One mode from each state is combined on a balanced beam splitter and the $\hat{X}$ and $\hat{P}$ quadrature of the outputs are measured. Success is conditioned on a value of the $\hat{X}$-outcome close to zero.} 
\label{fig:jonarep}
\end{figure}

The repeater protocol of Ref.~\cite{brask2010prl} consists of three steps,  (i) heralded entanglement generation based on sources of two-mode squeezed vacuum and SPD, (ii) growth of two-mode cat states from entangled single photons by means of homodyning, and (iii) entanglement swapping based on homodyning. The steps are outlined in \figref{fig:jonarep}.

In step (i) (see \figref{fig:jonarep}(i)), two sources produce two-mode squeezed vacuum states of the form
\begin{equation}
\ket{0,0} + \sqrt{p_{pair}} \ket{1,1} + O(p),
\end{equation}
where $p_{pair}$ is the probability to produce a photon pair. These sources can be realized using parametric downconversion crystals or ensembles of $\Lambda$-type atoms \cite{duan2001,simon2007}. One output mode from each source is read into a quantum memory while the remaining modes are sent to a balanced beam splitter positioned between the two sources. The beam-splitter outputs are measured and a single SPD click projects the two modes in the quantum memories into an entangled state $\ket{0,1}+\ket{1,0}$. The pair-production probability $p_{pair}$ (and hence the squeezing) needs to be small to ensure that the final state does not contain more than a single photon.

In step (ii) (see \figref{fig:jonarep}(ii)), two entangled single-photon states are combined on balanced beam splitters and the $\hat{X}$ quadratures of one output mode from each beam splitter are measured (a related procedure to perform distillation of continuous-variable entanglement was shown in Ref. \cite{eisert2003,eisert2004}). Whenever the sum of the measurement outcomes fulfill $|x_a + x_b| \leq \Delta$, for a certain acceptance interval $\Delta$, the state is kept.  The process can be iterated by combining two states resulting from successful growth at the previous level and repeating the procedure. In the limit of small $\Delta$, the final output resulting from this procedure approaches a non-locally squeezed two-mode cat state of the form
\begin{equation}
\label{eq:targetone}
\hat{S}_{+}(2)\ket{\gamma(0,\mu_{m}/\sqrt{2})},
\end{equation} 
where $\mu_{m}=\sqrt{2^{m}+1/2}$ and $m$ is the number of iterations. $\hat{S}_{+}(2)$ denotes non-local squeezing in the variance of $\hat{X}_{a}+\hat{X}_{b}$ by a factor of two. The squeezing operator has the general form $\hat{S}(\zeta)=\text{exp}\left(\frac{1}{2} \zeta^{*}\hat{a}^{2}-\frac{1}{2} \zeta\hat{a}^{\dagger 2}\right)$. The acceptance interval $\Delta$ determines the probability for successful growth, and hence the rate, as well as the fidelity of the output state with respect to the state in Eq. \eqref{eq:targetone}. Larger $\Delta$ corresponds to higher success probability but lower fidelity. The choice of $\Delta$ thus defines a tradeoff between the rate and the fidelity. In Ref.~\cite{brask2010prl}, $\Delta$ was fixed to take the same value for all $m$.
 
The final step (iii) (see \figref{fig:jonarep}(iii)) is entanglement swapping where neighboring entangled segments are combined to create longer segments. Two modes, one from each entangled pair, are combined on a balanced beam splitter and the $\hat{X}$ and $\hat{P}$ quadratures of the output modes are measured. Whenever $|x| \leq \delta$, $x$ being the outcome of the $\hat{X}$ measurement, the entanglement swapping is considered to be a success and the output state is kept. The process is iterated until entanglement is distributed over the total length, $L$ of the repeater. This is obtained by first dividing $L$ into $2^{n}$ segments of length $L_{0}=L/2^{n}$ over which entanglement is created. At each swap level, every two neighbouring segments are connected, such that the entanglement distance is doubled. After $n$ swap levels entanglement is distributed over the entire length $L$.  In the limit of small $\delta$, the state produced after $n$ swap levels approaches a locally squeezed two-mode cat state 
\begin{equation}
\label{eq:targettwo}
\ket{\psi_{ideal}}_{ab}=\hat{S}_{a}(\sqrt{2})\hat{S}_{b}(\sqrt{2})\ket{\gamma(\phi_{n},2^{-5/4}\mu_{m})}_{ab},
\end{equation}
where the phase $\phi_{n}$ depends on the $\hat{P}$-measurement outcomes from the previous levels. This state contain one ebit of entanglement and is used to quantify the performance of the repeater via the fidelity 
\begin{equation}
F_{prev} = \abs{\matrixel{\psi_{ideal}}{\hat{\rho}}{\psi_{ideal}}}^{2},
\end{equation} 
where $\hat{\rho}$ is the density matrix of the final output state of the repeater. As for the growth step (ii), there is a tradeoff between fidelity and rate through the acceptance parameter $\delta$. The upper limit to the success probability of entanglement swapping is $1/2$ for the simple procedure considered here but the success probability can be increased by using a more complicated procedure. For simplicity we will not consider this here. 
 
\section{The modified scheme}
\label{sec.newrep}

\begin{figure} [H]
\centering
\includegraphics[width=0.5\textwidth]{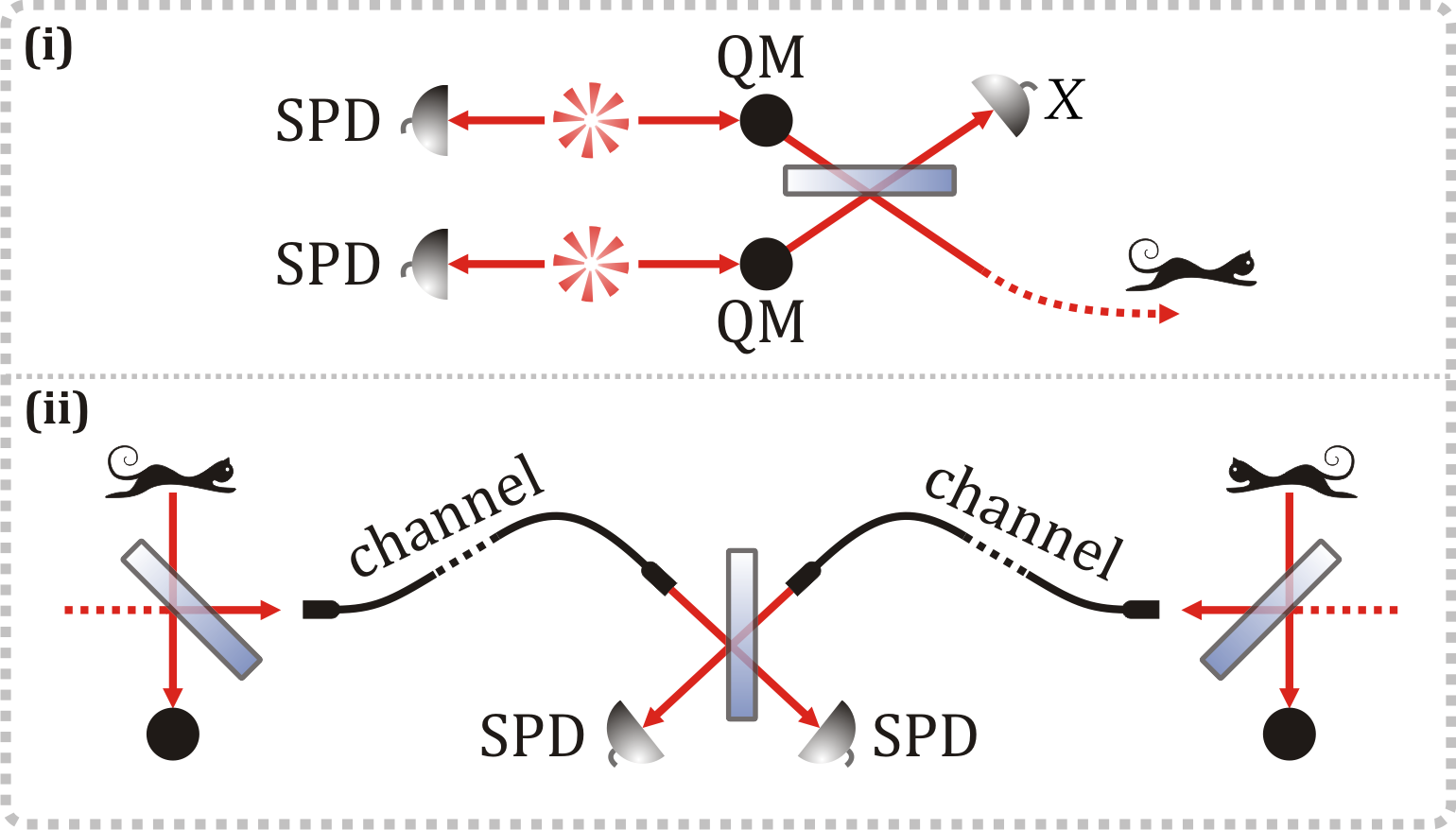}
\caption{(Colour online) Steps of the modified repeater. \textbf{(i)} Local growth of cat states. The output modes from two sources of two mode squeezed vacuum states are combined on a balanced beamsplitter and the $\hat{X}$ quadrature of one of the outputs is measured conditioned on a click in both SPD-detectors. Conditioned on the measurement outcome the resulting state is kept for further processing. \textbf{(ii)} Entanglement generation from single-mode cat states. Small parts are tapped off from two input cat states at two separate locations, and the remaining parts are stored in quantum memories (black dots). The fraction tapped off is controlled by the reflectivity $r$ of the local beam splitters. The reflected signals are transmitted to a central, balanced beam splitter, and the output ports are measured. Conditioned on a click in either of the detectors, the memories are prepared in an entangled state.}
\label{fig:newrep}
\end{figure}

To improve the rate of entanglement distribution, we interchange steps (i) and (ii) above, resulting in a new protocol based on local growth of single-mode cat states and subsequent non-local single-photon subtraction. The steps of the new protocol are sketched in \figref{fig:newrep}. Entanglement swapping is performed in the same manner as in step (iii) above.

\subsection{Growth of cat states}

The first step of the modified protocol is growth of states approximating squeezed single-mode cat states
\begin{equation}
\label{eq:sqcat}
\ket{\zeta_{m}}=\hat{S}(2)\frac{1}{\sqrt{N_{\mu_{m}}^{+}}}(\ket{\mu_{m}}+\ket{-\mu_{m}}) , 
\end{equation}
where $\hat{S}(2)$ denotes squeezing by a factor of two in the variance of the $\hat{X}$ quadrature, $\ket{\mu_{m}}$ is a coherent state with amplitude $\mu_{m}=\sqrt{2^{m}+1/2}$ and $N_{\mu_{m}}^{+}$ is a normalization constant. These states can be grown by a setup very similar to step (ii) of the original protocol, as explained in Ref.~\cite{brask2010prl}. The input states are single-mode one-photon states, generated by detecting one half of a two-mode squeezed state with small pair-production probability $p$. To understand the growth procedure, we consider the ideal limit where each source produces a pure single-photon state $\ket{1}$ with corresponding wave function for the $x$-quadrature
\begin{equation}
\psi_{0}(x)=\frac{\sqrt{2}}{\pi^{-1/4}}e^{-\frac{1}{2}x^{2}}x.
\end{equation}
The joint wave function before the beam splitter is $\psi_{0}(x)\psi_{0}(y)$. At the output of the balanced beam splitter, this is transformed into $\psi_{0}((x+y)/\sqrt{2})\psi_{0}((x-y)/\sqrt{2}) \varpropto e^{-\frac{1}{2}(x^{2}+y^{2})}(x^{2}-y^{2})$. Now mode $y$ is measured and the state is kept if $y_{0} \in [-\Delta,\Delta]$ where $y_{0}$ is the measurement outcome. Taking the limit $\Delta \to 0$ we find the output state $\psi_{1}(x) \varpropto e^{-\frac{1}{2}x^{2}}x^{2}$. Then the process is iterated with $\psi_1(x)$ as input. After $m$ iterations the output wave function becomes
\begin{equation} \label{eq:statem1} 
\psi_{m}(x) = \Gamma \left(2+\frac{1}{2} \right)^{-\frac{1}{2}}x^{2^{m}}e^{-\frac{1}{2}x^{2}}.
\end{equation}
The overlap of this state with the state in Eq. \eqref{eq:sqcat} exceeds 99\% for $m\ge2$ and approximate squeezed cat states can thus be grown this way.

As in the previous section, there will be a tradeoff between the fidelity and the rate controlled by $\Delta$. In Ref.~\cite{brask2010prl} $\Delta$ was kept fixed at the same value in every iteration but here we investigate the improvement by allowing different values of $\Delta$ for each $m$. To understand the possible improvement allowed by varying the interval we first analyze how the growth procedure works. The output wave function of the growth procedure (approximately $\psi_m(x)$ for small $\Delta$) is symmetric with two peaks; one at $x<0$ and one at $x>0$. Suppose that the measurement is performed in the symmetric output of the beam splitter with quadrature operator $\hat{X}_{+}=\hat{X}_{1}+\hat{X}_{2}$. In this mode the quadratures add. If the two peaks with positive $x$ are combined the measurement outcome will likely have a positive value. Similarly, combining the negative peaks leads to a negative outcome. These two possibilities are not desirable since the wave function in the antisymmetric mode $\hat{X}_{-}=\hat{X}_{1}-\hat{X}_{2}$ essentially will be a peak around zero, because in this mode the quadratures subtract. However, when a negative and a positive peak combine, the measurement outcome will be in the vicinity of zero. Since there are two paths leading to this result, corresponding to two different states in the antisymmetric output mode, the desired cat state is generated. The acceptance interval must be chosen such that one avoids outcomes resulting from the tail of the distribution coming from the combination of two positive or two negative peaks. The closer the peaks are to each other at the input, the smaller acceptance interval is allowed. As the growth process is iterated, the peaks become more separated and larger acceptance intervals can be chosen, resulting in a higher probability of success.

\begin{figure} [H]
\centering
\subfloat {\label{fig:m=2}\includegraphics[width=0.45\textwidth]{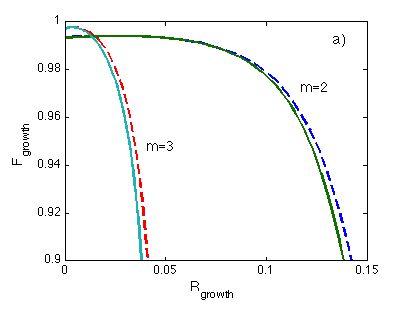}} \\
\subfloat{\label{fig:m=3}\includegraphics[width=0.45\textwidth]{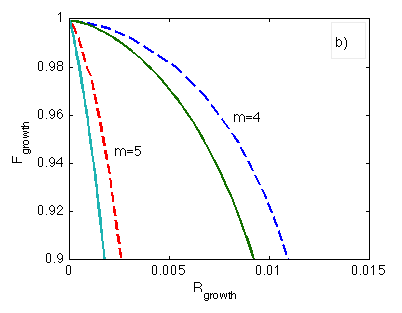}}
\caption{(Colour online) Optimized production rate of approximate squeezed cat states after (a) m=2,3 iterations (b) m=4,5 iterations. The dashed lines are the optimal curves and the solid curves are obtained for identical acceptance intervals in all iterations. The fidelity was calculated with the target state in Eq. \eqref{eq:sqcat} and the rate is given in units of the source repetition rate.}
\label{fig:optimcat}
\end{figure}

We have optimized the acceptance interval to achieve the highest possible probability for a fixed target fidelity
\begin{equation}
F_{growth}=\matrixel{\zeta_{m}}{\hat{\rho}_{m}}{\zeta_{m}}
\end{equation}
of the output state $\hat{\rho}_{m}$ of the growth procedure. We assume perfect one-photon states at the inputs, and calculate the fidelity and rate on a grid of values for each acceptance interval, $\Delta_{m}$ under the constraint that $\Delta_{m+1}\ge\Delta_{m}$. The optimization was made using Wigner functions, since these provide a natural description of mixed continuous-variable states and make it possible to compute the average output fidelity. Details are given in \appref{sec:appgrowing}. The rate, in units of the source repetition rate is approximated by
\begin{equation} \label{eq:rategrowapprox}
R_{growth}=\left(\frac{3}{2}\right)^{m-1}P_{1}P_{2}\ldots P_{m} ,
\end{equation}
where $P_{m}$ is the probability of successful growth in iteration $m$. This expression assumes that the outcomes of successful events can be stored while unsuccessful events are repeated until they succeed \cite{sangouard2011}. Note that by assuming that successful events are stored in quantum memories we avoid the usual exponential scaling with the number of conversion events (recall that the number of down conversions is $2^{m}$). The result of the optimization is shown in \figref{fig:optimcat} where we plot $F_{growth}$ against $R_{growth}$. Note that the fidelity does not reach unity as $R_{growth} \to 0$ since we take the fidelity with the approximate cat state in Eq. \eqref{eq:sqcat} and not the state in the ideal limit in Eq. \eqref{eq:statem1}. The calculation was restricted to $m\le5$ for runtime reasons.   

\figref{fig:optimcat} shows that the growth procedure is indeed improved by allowing for different acceptance intervals in every iteration. However the rate is not significantly improved for a small number of iterations $m$. Nonetheless Fig. \ref{fig:optimcat} indicates that the improvement will increase with $m$ since for larger $m$ the peaks in the input states become more separated. For an output fidelity of 0.9 the ratio of the modified rate to the previous rate is 1.03, 1.10, 1.21, and 1.53 for $m=2,3,4$, and 5 respectively. For very large $m$ the two peaks will be so far separated that we can choose an acceptance interval for which the success probability approaches $\frac{1}{2}$ without affecting the fidelity. 

\subsection{Connection of cat states}

The second step of the new protocol is to create entanglement by connecting the single-mode cat states from the first step. The method we employ was proposed by N.~Sangouard and coworkers in Ref.~\cite{sangouard2010} and the setup is shown in Fig.~\ref{fig:newrep}(ii).

A small part is subtracted from each input state by means of asymmetric beam splitters with low reflectivity $r$. The remaining parts of the states are stored and the reflected parts are transmitted to a central station. Here, the two signals are combined on a balanced beam splitter and the two output ports are measured with photodetectors. Successful entanglement generation is conditioned on a click in exactly one of the detectors. When a click is observed in the symmetric output port of the beam splitter, the quantum memories are projected into an entangled state approximating the two-mode cat $\ket{\gamma(0,\alpha \sqrt{1-r})}$. For a click at the antisymmetric output, the state is identical up to a local phase shift. The procedure can be understood easily in the ideal case where the inputs are exact cat states $\ket{\xi(0,\alpha)}$ and $r \to 0$. In this limit, a click heralds non-local subtraction of a single photon from the joint state of the memories. The memories are thereby projected into the (unnormalized) state
\begin{equation}
(\hat{a}\pm\hat{b})\ket{\xi(0,\alpha)}_{a}\ket{\xi(0,\alpha)}_{b},
\end{equation}
where $a$, $b$ label the output modes. Inserting the definition \eqnref{eq:singcat} and recalling that coherent states are eigenstates of the annihilation operators, we notice that the component of the wave function where the $a$ and $b$ mode have the opposite (same) phase e.g. $\ket{\alpha}_{a}\ket{-\alpha}_{b}$ ($\ket{\alpha}_{a}\ket{\alpha}_{b}$) vanish by interference for the plus (minus) combination. Therefore the resulting state is
\begin{equation}
\ket{\alpha}_{a}\ket{\pm\alpha}_{b}- \ket{-\alpha}_{a}\ket{\mp\alpha}_{b} .
\end{equation}
Comparing to Eq. \eqnref{eq:cat}, we see that up to a local phase shift this state is equal to $\ket{\gamma(\pi/2,\alpha)}$.

In practice, it is very hard to create genuine cat states. Therefore we shall use the approximate squeezed cat states from the previous step of the repeater protocol. We now examine the behaviour of these states under entanglement generation in the ideal limit $\Delta \to 0$, in which case they are given by Eq. \eqref{eq:statem1}, and taking again the limit $r \to 0$, we find
\begin{equation}
\label{eq:psim}
\begin{split}
\ket{\Psi_{m}}_{ab} & =(\hat{a} \pm \hat{b})\ket{\psi_{m}}_{a}\ket{\psi_{m}}_{b} , \\
& \propto \frac{1}{\sqrt{2}} (\ket{0_{m}}_{a}\ket{1_{m}}_{b}\pm\ket{1_{m}}_{a}\ket{0_{m}}_{b}) ,
\end{split}
\end{equation}
where 
\begin{equation}
\begin{split}
\braket{x}{0_{m}} & = \matrixel{x}{a}{\psi_{m}}=\Gamma(2^{m}-1/2)^{-1/2}x^{2^{m}-1}e^{-\frac{1}{2}x^{2}} , \\
\braket{x}{1_{m}} & =\psi_{m}(x)= \Gamma(2^{m}+1/2)^{-1/2}x^{2^{m}}e^{-\frac{1}{2}x^{2}} .
\end{split}
\end{equation}
Here, $\ket{1_{m}}$ is a superposition of even photon states and $\ket{0_{m}}$ is a superposition of odd photon states, and for $m\ge2$ they resemble squeezed even and odd single-mode cat states respectively i.e:
\begin{eqnarray}
\ket{1_{m}}\approx \hat{S}(2)\frac{1}{\sqrt{N_{\mu_{m}}^{+}}}(\ket{\mu_{m}}+\ket{-\mu_{m}}) \label{eq:approx1}\\
\ket{0_{m}}\approx \hat{S}(2)\frac{1}{\sqrt{N_{\tilde{\mu}_{m}}^{-}}}(\ket{\tilde{\mu}_{m}}-\ket{-\tilde{\mu}_{m}}) \label{eq:approx2}
\end{eqnarray}
where $\tilde{\mu}_{m}=\sqrt{2^{m}-1/2}$, $\mu_{m}=\sqrt{2^{m}+1/2}$ and $N_{\mu_{m}}^{+},N_{\tilde{\mu}_{m}}^{-}$ are normalization constants. For $m\ge2$ the fidelities between $\ket{1_{m}},\ket{0_{m}}$ and the respective cat states are both $\ge99\%$. 

The state $\ket{\Psi_{m}}$ contains one ebit of entanglement and is obtained in the low-rate limit of small acceptance intervals during growth and small reflectance during connection. $\ket{\Psi_{m}}$ however deviates from the squeezed two-mode cat state that was shown to be useful for entanglement swapping in Ref.~\cite{brask2010prl}. For $m=2$ the overlap with a locally squeezed two-mode cat state of the form used in Ref.~\cite{brask2010prl}
 \begin{equation} \label{eq:sqtwocat}
\hat{S}(2)_{a}\hat{S}(2)_{b}\ket{\gamma(0,2^{m/2})}_{ab} ,
\end{equation} 
is $96\%$ and for $m=3$ it is $97\%$. As we will see, this discrepancy has a detrimental effect on the overall performance of the repeater. This could be avoided by unsqueezing the approximate squeezed cat states going into the entanglement connection. Such unsqueezing operations may however be technically demanding, and we prefer not to include them here. We therefore consider the simplest situation where we directly connect the states generated in the first step. Alternatively, the problem could be mitigated by increasing $m$. E.g. for $m\ge5$ we get an overlap of $99\%$ with the state in Eq. \eqnref{eq:sqtwocat}. However, going to such high $m$ would also be very demanding in practice.

\begin{figure} [H]
\centering
\includegraphics[width=0.39\textwidth]{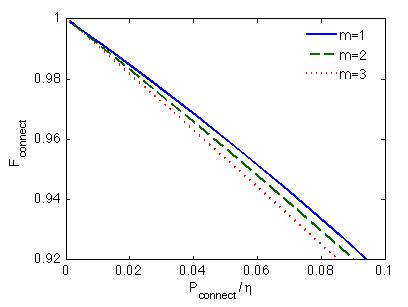}
\caption{(Colour online) The fidelity of the connected state with respect to the state $\ket{\Psi_{m}}$ plotted against the rescaled probability of a successful connection. We have assumed the input states to be of the form in Eq. \eqref{eq:statem1}. }
\label{fig:rdepend}
\end{figure}

Below we analyse the full repeater protocol, including entanglement swapping. Before proceeding we first examine the performance of the connection step itself. We compute the output state for finite $r$ and lossy transmission channels. Loss is modelled by fictitious beam splitters of transmittivity $\eta$, such that the probability for a photon to get lost on the way to the central station is $1-\eta$. We assume that the photodetectors do not resolve the photon number. Details of the calculation are given in \appref{sec:appconnect}. First we study the output fidelity, $F_{connect}$ of the connected state with respect to $\ket{\Psi_{m}}$ as a function of the reflectivity, $r$ of the first two beamsplitters. For this purpose we simulate the connection of states of the form in Eq. \eqref{eq:statem1} for a fixed number of iterations ($m$). We restrict the simulations to small $r$ since this is the relevant regime of the repeater. This implies that the probability of a successful connection is $P_{connect}\approx P_{c,noloss}(r)\eta$ where $P_{c,noloss}(r)$ is independent of the losses in the optical fibers. The results of the simulations are shown in \figref{fig:rdepend}.

\begin{figure} [H]
\centering
\includegraphics[width=0.4\textwidth]{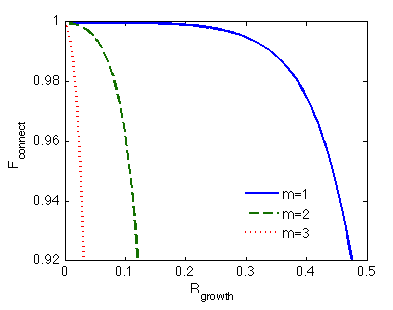}
\caption{(Colour online) Influence of finite acceptance intervals in the growth on the state after connection. The fidelity of the connected state is w.r.t. the state $\ket{\Psi_{m}}$ and $\vec{\Delta}$ is represented through $R_{growth}$. Here we have neglected losses in the optical fibers and $R_{growth}$ is in units of the rate at which the one-photon input states for the growth can be provided. }
\label{fig:deltadepend}
\end{figure}

 \figref{fig:rdepend} shows that the fidelity depends linearly on $P_{connect}/\eta$ in the limit of small $r$. Furthermore the rate of the connection step for a fixed distance is more or less independent of $m$ for small $r$. These results can be understood by noting that the connection fails if a second photon is tapped off at the beam splitters. The probability for this to happen conditioned on at least a single photon being tapped off is $\sim P_{connect}/\eta$ regardless of $m$. \\
 The second parameter to consider in the connection step is the vector of acceptance intervals for the growth step, $\vec{\Delta}$, which determines the fidelity of the input states with respect to $\ket{\psi_{m}}$. To determine the effect of finite acceptance intervals in the growth procedure on the state after connection, we simulate the connection step for different $\vec{\Delta}$, taking the limit of $r \to 0$ and $\eta \to 0$. We take $\vec{\Delta}$ to be the vectors giving the optimal fidelity for a given rate $R_{growth}$ in \figref{fig:optimcat}. The result of the simulations is shown in \figref{fig:deltadepend}. 
 
\figref{fig:deltadepend} shows the same kind of behavior as \figref{fig:optimcat} taking into account that the fidelity in \figref{fig:optimcat} is w.r.t. the squeezed one mode cat state in Eq. \eqref{eq:sqcat}, i.e. as opposed to \figref{fig:optimcat} the fidelity approaches unity. For optimizing the performance of the full repeater it is advantageous to have an analytical understanding of the entanglement generation. We have therefore fitted the graphs to functions of the form $F_{connect}=1-c*e^{d*R_{growth}}$. The details of the fits are shown in \appref{app:supma} and \tabref{tab:table2}.
 \figref{fig:rdepend} and \figref{fig:deltadepend} show that the highest rate of entanglement generation is obtained for $m=1$ but as we will see below we need to go to higher $m$ for the swapping procedure to function. 
 
\subsection{Entanglement swapping}

\begin{figure} 
\centering
\subfloat{\label{fig:fidel11}\includegraphics[width=0.45\textwidth]{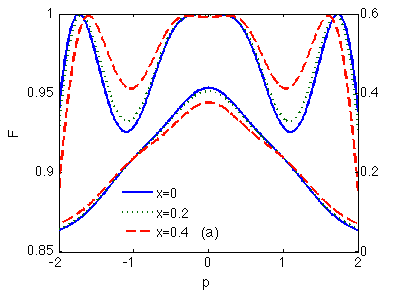}} \\
\subfloat {\label{fig:fidel12}\includegraphics[width=0.45\textwidth]{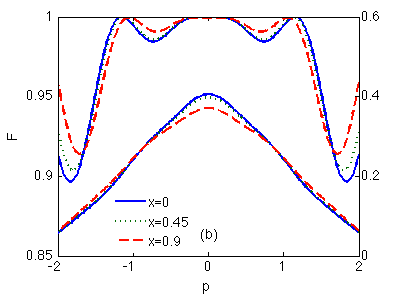}} \\
\subfloat {\label{fig:fidel13}\includegraphics[width=0.45\textwidth]{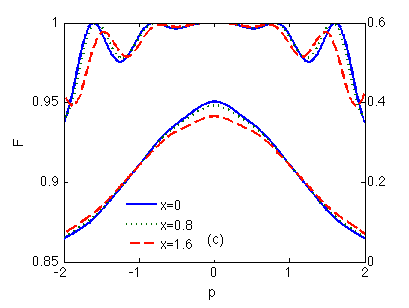}} 
\caption{(Colour online) Upper curves (left axis): The fidelity after entanglement swapping as a function of the $\hat{P}$-outcome for (a) $m=1$, (b) $m=2$ and (c) $m=3$, and various values of the $\hat{X}$-outcome. The fidelity is $F(\hat{\rho}_\Psi)$  where $\hat{\rho}_{\Psi}$ results from swapping two copies of $\ket{\Psi_{m}}$. Lower curves (right axis): The corresponding probability distributions of $p$ for each $\hat{X}$-outcome.}
\label{fig:fidelone}
\end{figure}

The final step of our altered repeater is to merge entangled segments via entanglement swapping. The method is the same as in the protocol of Ref.~\cite{brask2010prl} and is illustrated in \figref{fig:jonarep}(iii). Two modes at the same location from two entangled pairs are connected on a balanced beam splitter and the $\hat{X}$ and $\hat{P}$ quadratures are subsequently measured. Whether the swap attempt was successful is conditioned on the outcome of the $\hat{X}$ measurement. When swapping two states of the form \eqnref{eq:sqtwocat}, the wave functions of the states have two peaks; one at $x>0$ and one at $x<0$. Thus, following similar arguments as for the growth procedure, there are two paths leading to outcomes in the vicinity of zero, $|x| \leq \delta$. Measuring the plus combination there is one from the first mode having a positive value of $x$ combined with a negative value from the second mode and vice versa. If $x \sim 0$, the two remaining quantum memories are projected into an entangled state of the form in Eq. \eqnref{eq:sqtwocat} with a phase determined by the outcome of the $\hat{P}$ measurement. The entangled states produced in the connection step are however not exactly of the ideal form in Eq. \eqnref{eq:sqtwocat}. Therefore we need to investigate how the swapping performs with the actual states generated by our protocol. To this end we first identify the entangled state that most closely resembles the result of swapping after ideal growth and connection by swapping states of the form in Eq. \eqref{eq:psim}. Swapping two copies of $\ket{\Psi_{m}}$ using the approximations \eqnref{eq:approx1}, \eqnref{eq:approx2} to determine how the two modes gets mixed we find
\begin{align}
\label{eq:targetswap}
\ket{\Phi_m} = & A\ket{0_{m}}\ket{0_{m}}-A^{*}\ket{1_{m}}\ket{1_{m}} + \nonumber \\
& C\ket{1_{m}}\ket{0_{m}}+C^{*}\ket{0_{m}}\ket{1_{m}} ,
\end{align}  
where the coefficients depend on the measurement outcomes of the $\hat{X}$ and $\hat{P}$ measurements in both the current and previous swap levels (see \appref{sec:appswap} for details). This state contains one ebit of entanglement, and we will use it as our target state when evaluating the performance of the repeater. That is, we measure the quality of a final state $\hat{\rho}$ produced by the repeater by the fidelity
\begin{equation}
\label{eq:finalfid}
F(\hat{\rho}) = \abs{\matrixel{\Phi_m}{\hat{\rho}}{\Phi_m}}^{2} .
\end{equation} 
The approximate form in Eq. \eqref{eq:targetswap} is however only obtained in the limit of large $m$. For finite $m$, even the state $\ket{\Psi_{m}}$, obtained in the limit of ideal growth and connection, will produce less than one ebit of entanglement. To quantify this, we examine the dependence of $F$ on the outcome of the $\hat{P}$ measurement. This behavior is shown in \figref{fig:fidelone} where we plot $F$ against the $\hat{P}$-outcome for different values of $m$. For small values of $m$, there is a strong dependence. However as $m$ increases the $\hat{P}$ dependence decreases because the states begin to resemble locally squeezed two mode cat states, which are insensitive to the $\hat{P}$-outcome when swapped. 
\noindent
The probability of a successful swap is determined by the acceptance interval $\delta$ for the outcome of the $\hat{X}$-measurement and has an upper bound of $1/2$, which is approached for high $m$ as in the protocol of Ref.~\cite{brask2010prl}. Near-deterministic swapping can in principle be achieved following the method of Ref.~\cite{brask2010prl} using auxiliary single-mode cat states but we will not consider this possibility here.  \\

The strong $\hat{P}$ dependence for small $m$ in the fidelity of the swapped state was not seen in the original hybrid repeater \cite{brask2010prl} where the the outcome of  the $\hat{P}$ measurement merely resulted in an overall phase in the swapped state. As a consequence the states produced in the connection step of the altered repeater do not swap as well as those in the original repeater for the same number of iterations $m$. For long distances a large number of swap levels is needed. One therefore needs to go to higher $m$ in the altered repeater as compared to the original repeater to reach a given output fidelity of the distributed state.  

\section{Performance}
\label{sec.perform}

\begin{figure} [H]
\centering
\includegraphics[width=0.5\textwidth]{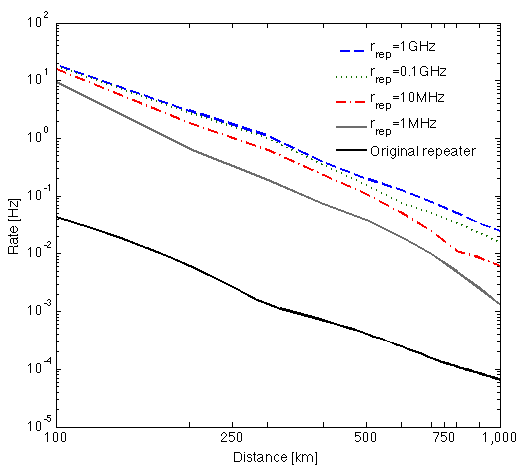}
\caption{(Colour online) The optimal rates of the present and previous repeater protocols for different values of $r_{rep}$. The protocols are optimized over the parameters listed in Table 1, under the constraint $F\geq 80\%$. The altered repeater performs significantly better than the previous protocol even for $r_{rep} = 1$MHz.}
\label{fig:optimrate2}
\end{figure}

The full repeater protocol is the nested collection of the three steps described above i.e.~growth of cat states, connection, and entanglement swapping. To quantify the performance of the repeater we use the fidelity $F$, as given in Eq. \eqref{eq:finalfid}, and the production rate for the final entangled states. We set a target value of $F \geq 80\%$ and make a numerical optimization of the rate as a function of distance by simulating the repeater for different values of the control parameters at each step. The relevant parameters are given in \tabref{tab:table2}. We perform a full optimization over all the parameters in \tabref{tab:table2}, under the constraint that the final state should have a minimum fidelity $F\geq 80\%$. We do this optimization for each distance and for each value of the local repetition rate. For the simulation, we assume perfect quantum memories, perfect homodyning, and a SPD efficiency $\eta_{spd}=50\%$. For a repeater of total length $L$ and $n$ swap levels, the distance between the stations is $L_0=L/2^n$ and the transmission efficiency incurred in the entanglement generation step is $e^{-L_0/2L_{att}}$, where $L_{att}$ is the attenuation length of the channels. The total efficiency incurred is thus $\eta = \eta_{spd}e^{-L_0/2L_{att}}$. We assume $L_{att} = 20\text{km}$ corresponding to optical fibers at telecom wavelengths. The time needed for classical communication during entanglement generation is given by $L_0/c$, where $c$ is the speed of light in the channels. We assume $c=2 \cdot 10^{5}\text{km/s}$. The time required for local operations (measurements and memory operations) is assumed to be negligible compared to the classical communication time, such that the characteristic rates in the protocol are $c/L_0$ and the source rate for the two-mode squeezing sources $r_{rep}$. The latter is taken to be the repetition rate of a single two mode squeezing source, i.e. the rate at which down conversion is attempted in a single crystal. The optimal pair production probability $p_{pair}$ is found in the numerical optimization of the rate of the repeater for a given $r_{rep}$. The effect of two-photon contributions in the input states is treated by perturbation in the pair-production probability $p_{pair}$, as in Ref.~\cite{brask2010prl}. $r_{rep}$ determines the rate of the growth, which is the first step of the protocol and thus has a large effect on the overall rate of the repeater.  For runtime reasons, we have restricted the number of growth steps to $m \le 3$ and the number of swap levels to $n \le 4$. 
\\
For simulating both the growth and the connection step of the repeater we use Wigner functions to obtain the average output fidelity for a given set of values of the control parameters (see \appref{sec:appgrowing} and \appref{sec:appconnect}). However this is not possible when simulating the entanglement swapping since the target state depends on the outcomes of the $\hat{X}$ and $\hat{P}$ measurements. To obtain an average fidelity of the entanglement step we therefore pick the measurement outcomes according to the probability distributions of $\hat{X}$ and $\hat{P}$ and calculate the fidelity of the resulting state. We repeat this procedure 100 times for each swap level and calculate the average output fidelity. This gives a standard deviation of the mean of the fidelity of about 1\%. \\ \\
When performing the numerical optimization of the rate we calculate the fidelity of the distributed state and the rate on a grid of values for all the control parameters. The parameters affecting the performance of the repeater are summarized in \tabref{tab:table2}. In order to pinpoint the relevant parameter regime we use the fits listed in \tabref{tab:table2} to make an analytical approximation of how the fidelity depends on the different parameters. We use this approximation to optimize the rate using the method of Lagrange multipliers to find the optimal rate for a target fidelity of 80\%. The resulting values of the control parameters is then used to make a grid of values for the numerical optimization around the analytical results. Finally we pick the grid point with the highest rate where $F \geq 80\%$. The optimal rate as a function of distance is shown in \figref{fig:optimrate2} for different values of $r_{rep}$. 

\begin{table*} 
\begin{tabular}{| l | p{4cm} | p{5cm} | p{6cm} | }
\hline
Parameter & Description & Effect & Fidelity-fit \\ \hline
$p_{pair}$ & Pair-production probability of the sources of two-mode squeezed vacuum states. & Small $p_{pair}$ $\to$ low production rate of input states. \newline Large $p_{pair}$ $\to$ large two-photon component. &$F=(1-\tau\cdot p_{pair})F_{1}+\tau\cdot p_{pair}\cdot F_{2}$ \\ \hline 
$\vec{\Delta}$ & Vector of acceptance intervals in the growth procedure. & Large acceptance intervals $\to$ high growth rate. \newline Small acceptance intervals $\to$ high fidelity of the one-mode states state in \eqref{eq:statem1}. & $F=1-\tilde{c}_{n,m}e^{\tilde{d}_{n,m}R_{growth}}$ \\ \hline
$m$ & Number of iterations in the growth step. & High $m$ $\to$ low growth rate. \newline Low $m$ $\to$ poor swapping states. & $F=\tilde{i}_{n} + \tilde{j}_{n}\cdot m^{2} +\tilde{k}_{n}\cdot m$  \\ \hline 
 $r$ & Reflectivity of the first two beam splitters in the connection step. & Large $r$ $\to$ high connection rate. \newline Small $r$ $\to$ high fidelity with the state \eqref{eq:psim}. & $F=1-\tilde{{a}}_{n,m}(\frac{P_{connect}}{\eta})^{2}-\tilde{b}_{n,m}\frac{P_{connect}}{\eta}$ \\ \hline   
$\delta$ & Acceptance interval in the swapping procedure. & $\delta$ determines the  probability of a successful swap and the fidelity of the output state. & $F=\tilde{e}_{n,m}e^{\tilde{f}_{n,m}\delta}+\tilde{g}_{n,m}e^{\tilde{h}_{n,m}\delta}, \quad m\le2$ \newline $F=\tilde{e}_{n,3}+\tilde{f}_{n,3}\cdot \delta \qquad \qquad \quad, \quad  m=3$  \\ \hline 
$n$ & Number of swap levels. & $n$ determines the classical communication time ($L_{0}/c$) between the stations in the elementary segments and hence the loss in the fibers during connection. & $F=1-\tilde{l}_{m}\cdot n^{2}$ \\
 \hline 
\end{tabular}
\caption{Parameters considered in the  numerical optimization of the repeater. The last column is a functional fit of how the fidelity \eqnref{eq:finalfid} depends on the parameter when the other parameters assume their ideal values i.e.~$p_{pair} \to 0$, $\vec{\Delta} \to 0$, $r \to 0$, $\delta \to 0$. $r$ is represented through $P_{connect}$ in the fit where $P_{connect}$ is the probability of a successful connection and $\vec{\Delta}$ is represented through $R_{growth}$. The fits for $R_{growth}, P_{connect}$ and $\delta$ are made for a specific choice of $n$ and $m$. See \appref{app:supma} for details on the matrices containing $\tilde{a}_{n,m}..\tilde{h}_{nm}$ and the vectors containing $\tilde{i}_{n}..\tilde{k}_{n}$ and $\tilde{l}_{m}$. The expression for the fidelity's dependence on $p_{pair}$ is calculated by perturbation in $p_{pair}$. $F_{1}$ is the fidelity with pure one-photon input states for a given set of parameters and $F_{2}$ is the fidelity for the same set of parameters but with one of the input states being a two-photon state. $\tau=\frac{f_{2}}{4}2^{m+n+1}$ where $f_{2}$ is a factor that accounts for the different acceptance probabilities for a one-photon state and a two-photon state in the repeater.} 
\label{tab:table2}
\end{table*}

Naturally the rate of the altered repeater is very dependent on the source repetition rate. With a fast local repetition rate cat states can be grown rapidly thus removing a time consuming step of the original repeater where this was done non-locally. Assuming an experimentally accessible repetition rate of 1MHz, the present protocol achieves a rate of $\sim 0.08$ pairs/min at $L=1000$km while the rate of the previous protocol for the same distance and target fidelity is $\sim 0.004$ pairs/min (note error in \cite{brask2010prl}). The altered repeater thus gives a significant increase in the rate. For $r_{rep} = 1$GHz the task of storing the signals in quantum memories will be challenging but a rate of $\sim 1.5$ pairs/min would in this case be reachable within the above assumptions. The ratio of the rate of the modified to that of the original repeater decreases as a function of the distance. This is because the states produced in the modified protocol are less robust to the swapping procedure than the states produced in the original repeater. When the distance increases the number of swap levels increase, which results in a decrease of the ratio of the rates for a fixed fidelity of the distributed state.

\section{Conclusion}
\label{sec.conclusion}

We have modified the quantum repeater protocol of Ref. \cite{brask2010prl} to improve the entanglement distribution rate. By interchanging the order of entanglement generation and growth of cat states, we have made the latter a local, hence faster, process, thus increasing the rate if local operations can be done rapidly. Furthermore, we have optimised the growth protocol. For entanglement generation, we have incorporated the method for connecting cat states of Ref.~\cite{sangouard2010}. We have performed a numerical simulation of our protocol, confirming that it does indeed lead to an increased rate. The final rate depends on the repetition rate of the two-mode squeezing sources at the base level of the protocol. For a moderate repetition rate of 1MHz, our protocol is 20 times faster than the repeater considered in Ref.~\cite{brask2010prl}, achieving a rate of $\sim 0.08$ pairs/min over 1000km. This rate is comparable to the best proposed atomic-ensemble based repeaters for similar detection efficiencies (taking into account that we have optimized for a final fidelity of 80\%) \cite{brask2010pra}. Working with discrete variables requires SPD efficiencies of $\sim 90\%$ to obtain similar rates or complicated swapping procedures \cite{sangouard2008}. Much higher source repetition rates than 1 MHz are plausible with parametric down conversion in nonlinear crystals, but compatible quantum memories operating at such high frequencies may be very difficult to implement. For quantum repeaters of this kind the most feasible quantum memories are currently those based on atomic ensembles. The high optical depth of  a dense ensemble of cold atoms enables a strong coupling even for a few photons and this can provide an increase of the bandwidth scaling as $\gamma d$, with $\gamma$ being the decay rate and $d$ the optical depth. For a sufficiently high $d$ the bandwidth may enable high repetition rates \cite{gorshkov2007}. Progress along this line was recently reported in Ref. \cite{reim2010}, which showed memory operations with pulses of spectral bandwidth exceeding 1 GHz. The storage-and-retrieval fidelities currently achievable are far from the perfect case assumed in the present analysis \cite{jensen2011}. The efficiency of an atomic ensemble memory can however in principle be made close to 100\% \cite{simon2010}. Since the modified repeater does not operate with bigger cat states than the original repeater we do not expect different scaling of the two when including inefficient quantum memories. Furthermore since the modified repeater operates faster than the original repeater the states do not need to be stored for as long a time, and the effects from decoherence will thus be smaller \cite{collins2007}. Thus we expect the improvement of the present protocol over the previous protocol to persist with at least the same factor even with non-ideal memories. It would be an interesting extension of our work to include non-ideal memories in the simulations, giving a more realistic calculation of the distribution rates but this is outside the scope of this article. 
 
\begin{acknowledgements}
We thank Nicolas Sangouard for useful discussions and we gratefully acknowledge the support of the Lundbeck Foundation, the Carlsberg Foundation, QUANTOP  and the Danish National Research Foundation.
\end{acknowledgements}

\appendix

\section{Growth of cat states}
\label{sec:appgrowing}

In this appendix we describe how the growth procedure transforms the Wigner function of the input states. In general we write the Wigner function as
\begin{equation}\label{eq:inone}
W_{m}(x,p)=\sum_{i=0}^{2^{m+1}}\sum_{j=0}^{2^{m+1}}w_{ij} x^{i}p^{j}e^{-(x^{2}+p^{2})}.
\end{equation} 
For a one photon state we have $m=0$ and the matrix containing $w_{ij}$ is
\begin{equation} \label{eq:wignone}
\mathbf{w}=\left(
\begin{array}{ccc}
-\frac{1}{\pi} & 0 & \frac{2}{\pi} \\
0 & 0 & 0 \\
\frac{2}{\pi} & 0 & 0 
\end{array} \right).
\end{equation}
The effect of the growth procedure in this representation is to change the size and elements of the matrix $\mathbf{w}$ along with the upper limit of the summations. The combination of two states of the form in Eq. \eqref{eq:inone} with variables $x,p$ and $x',p'$  on a balanced beam splitter is described by the transformations
\begin{equation}
\begin{split}
x  & \to \frac{1}{\sqrt{2}}\left(x+x' \right), \qquad p \to \frac{1}{\sqrt{2}}\left(p+p' \right) \\
x' & \to \frac{1}{\sqrt{2}}\left(x-x' \right), \qquad p' \to \frac{1}{\sqrt{2}}\left(p-p' \right).
\end{split} 
\end{equation}
Thus the state before the $\hat{X}$ measurement is
\begin{align}
\label{eq:mod1}
W'_{m+1}(x,p,x',p') = & \nonumber W_{m}\left(\frac{x+x'}{\sqrt{2}},\frac{p+p'}{\sqrt{2}}\right) \times \nonumber \\
& W_{m}\left(\frac{x-x'}{\sqrt{2}},\frac{p-p'}{\sqrt{2}}\right).
\end{align}
Using the identity $(a+b)^{i}=\sum_{s}^{i} \binom{i}{s}a^{s}b^{i-s}$ and collecting powers of $x$ and $p$ we can write $W'_{m+1}$ in the form:
\begin{align}
& W'_{m+1}(x,p,x',p') = \sum_{\{i,i'\}=0}^{2^{m+1}}\sum_{\{j,j'\}=0}^{2^{m+1}} \nonumber \\
& \sum_{k=0}^{i+i'} \sum_{s'=s_{\text{min}}}^{s_{\text{max}}} \binom{i}{k-s'} \binom{i'}{s'}(-1)^{i'-s'}x'^{i+i'-k}x^{k}  \nonumber \\ 
& \sum_{l=0}^{j+j'} \sum_{t'=t_{\text{min}}}^{t_{\text{max}}} \binom{j}{l-t'} \binom{j'}{t'} (-1)^{j'-t'}p'^{j+j'-l}p^{l}  \nonumber \\
& e^{-(x^{2}+x'^{2}+p^{2}+p'^{2})}w_{ij}w_{i'j'} ,
\end{align}
where 
\begin{equation}
\begin{split}
s_{\text{min}} & =\text{max}(0,k-i), \quad s_{\text{max}}=\text{min}(i',k) , \\
t_{\text{min}} & =\text{max}(0,l-j), \quad t_{\text{max}}=\text{min}(j',l) .
\end{split}
\end{equation}
The unnormalized average output after measuring $x'\in [-\Delta,\Delta]$ is found by integrating over momentum and position
\begin{equation}
\!\!\! \int \limits_{-\infty}^{\infty}\!\!\! \mathrm{d}p'\!\!\! \int \limits_{-\Delta}^{\Delta}\!\!\! \mathrm{d}x'W'_{m+1}(x,p,x',p') .
\end{equation}
After carrying out the integrals, we can write the unnormalized state after the growth procedure as
\begin{eqnarray} \label{eq:form2}
\tilde{W}_{m+1}(x,p)=\sum_{k=0}^{2^{m+2}}\sum_{l=0}^{2^{m+2}}\tilde{w}_{kl}x^{k}p^{l}e^{-(x^{2}+p^{2})} ,
\end{eqnarray}
with
\begin{align} \label{eq:wigit}
\tilde{w}_{kl} & = \sum_{\{i,i'\}=0}^{2^{m+1}}\sum_{\{j,j'\}=0}^{2^{m+1}}2^{-(i+i'+j+j')/2}(-1)^{i+i'+j+j'-k-l} \nonumber \\
& w_{ij}w_{i'j'}\kappa_{k}^{ii'}(\Delta)\kappa_{l}^{jj'}(\infty) ,
\end{align}
and
\begin{equation} \label{eq:matrixel1}
\kappa_{k}^{ii'}(t)= 
\begin{cases}
0 & \text{if }r<0 , \\
\sum \binom{i}{k-s'}\binom{i'}{s'}\int \limits_{-t}^{t}\mathrm{d}xe^{-x^{2}}x^{r} & \text{if } 0<r ,
\end{cases}
\end{equation}
where $r=i+i'-k$ and $\sum=\sum_{s'=s_{\text{min}}}^{s'=s_{\text{max}}}$. 
We have thus found a simple description for the Wigner function after a step of the growth procedure as function of the input Wigner function. To find the state after $m$ steps we start with the matrix in Eq. \eqref{eq:wignone} and iterate \eqref{eq:wigit} $m$ times. 

\section{Connecting Wigner functions}
\label{sec:appconnect}

In this appendix, we describe the connection step in terms of Wigner functions. The state before the two asymmetric beam splitters with reflectivity $r$ in the connection step is the product of the Wigner functions generated in step one of the repeater and two vacuum states
\begin{equation}
W_{m}(x,p)W_{m}(y,q)W_{vac}(x',p')W_{vac}(y',q').
\end{equation}
Here $W_{m}(-,-)$ has the form \eqref{eq:inone} and $W_{vac}(x,p)=\frac{1}{\pi}e^{-\frac{1}{2}(x^{2}+p^{2})}$.

The modes described by $(x,x',p,p')$ are on the left (location A) and the modes $(y,y',q,q')$ on the right (location B), (see \figref{fig:newrep}(ii)). Before the central station it is only necessary to focus on the modes described by $(x,x',p,p')$. Parametrising $\sin(\theta_r) = \sqrt{r}$, the action of the first beam splitter is 
\begin{align}
x  & \to \cos(\theta_r)x + \sin(\theta_r)x' , \nonumber \\
x' & \to \cos(\theta_r)x' - \sin(\theta_r)x ,
\end{align}
and the corresponding transformations on the momentum variables. This results in the state 
\begin{align}
& W_{a1}(x,x',p,p') = \qquad \qquad \qquad \qquad \qquad \quad \nonumber \\
& W_{m}(\cos(\theta_r) x + \sin(\theta_r) x' , \cos(\theta_r) p + \sin(\theta_r) p') \times \nonumber \\ 
& W_{vac}(\cos(\theta_r) x' - \sin(\theta_r) x , \cos(\theta_r)p' - \sin(\theta_r)p) . 
\end{align}
An additional beam splitter describing losses in the optical fibers mixes $x'$ and $p'$ with the vacuum mode described by $x''$ and $p''$. We parametrise the loss by $\sqrt{\eta} = \cos(\theta_{l})$. 
\begin{align}
x' & \to \cos(\theta_{l})x'+\sin(\theta_{l})x'' , \nonumber \\
x'' & \to \cos(\theta_{l})x''-\sin(\theta_{l})x' ,  
\end{align}
and the corresponding transformations on the momentum variables. The number of photons that are lost is not known and consequently we trace over $x''$ and $y''$. This produces the unnormalized state:
\begin{align}
& W_{a2}(x,x',p,p')=\int \limits_{-\infty}^{\infty}\!\mathrm{d}x'' \int \limits_{-\infty}^{\infty}\!\mathrm{d}y'' \nonumber \\
& W_{a1}(x,\cos(\theta_{l})x'\!\!\!+\sin(\theta_{l})x'',p,\cos(\theta_{l})p'\!\!\!+\sin(\theta_{l})p'') \times \nonumber \\
& W_{vac}(\cos(\theta_{l})x''\!\!\!-\sin(\theta_{l})x',\cos(\theta_{l})p''\!\!\!-\sin(\theta_{l})p') .
\end{align}
The modes described by $(y,y',q,q')$ is brought to the central beam splitter in the same manner producing the state $W_{b2}(y,y',q,q')$.The action of the central beam splitter is
\begin{equation}
x' \to \frac{x'+y'}{\sqrt{2}}, \quad y' \to \frac{x'-y'}{\sqrt{2}} ,
\end{equation}
and the corresponding transformations on the momentum variables. Assuming that one output mode only contains vacuum and the other contains anything but vacuum, the subsequent state is projected onto
\begin{equation}
W_{vac}(y',q')(1-W_{vac}(x',p'))
\end{equation}
Consequently the state in the quantum memories after the connection is
\begin{align}
& W_{ab}(x,y,p,q) = \frac{1}{N}\int \limits_{-\infty}^{\infty}\!\!\!\mathrm{d}x'\!\!\!\int \limits_{-\infty}^{\infty}\!\!\!\mathrm{d}y'\!\!\!\int \limits_{-\infty}^{\infty}\!\!\!\mathrm{d}p'\!\!\!\int \limits_{-\infty}^{\infty}\!\!\!\mathrm{d}q'\nonumber \\
& W_{vac}(y',q')(1-W_{vac}(x',p')) \times \nonumber \\
& W_{a2}(x,(x'+y')/\sqrt{2},p,(p'+q')/\sqrt{2}) \times \nonumber \\
& W_{b2}(y,(x'-y')/\sqrt{2},q,(q'-p')/\sqrt{2}) ,
\end{align}
where $N$ is the normalization constant.  
After the integration the resulting Wigner function can be written in the form
\begin{align}
\label{eq:form3}
& W_{ab}(x,y,q,p)=\qquad \qquad \qquad \qquad \qquad \nonumber \\
& \sum_{\{s,t,k,l\}=0}^{2^{m+1}}w_{stkl}x^{k}p^{l}y^{s}q^{t}e^{-x^{2}-p^{2}-y^{2}-q^{2}}.
\end{align}
This can be seen by writing $W_{a2}$ and $W_{b2}$ in the form of \eqref{eq:form2} and evaluating the integrals using the identity $(a+b)^{i}=\sum_{s=0}^{i}\binom{i}{s}a^{i}b^{i-s}$ as in \appref{sec:appgrowing}. The expression for $w_{stkl}$ is rather lengthy and we shall not reproduce it here. It can, however be implemented numerically and thus provide an efficient description of the connection step. 

\section{Target state of swapping}
\label{sec:appswap}

In this appendix we outline the calculations leading to Eq. \eqref{eq:targetswap} and give the expressions for the constants that appear in that equation.

We consider the swapping of two states of the type $\ket{\Psi_{m}}$ given in \eqref{eq:psim}. The state before the swap is thus
\begin{align}
\ket{\Psi_{m}}_{ab}\ket{\Psi_{m}}_{a'b'} \propto & (c_{1}\ket{0_{m}}\ket{1_{m}}+e_{1}\ket{1_{m}}\ket{0_{m}})_{ab} \times \nonumber \\ 
& (c_{2}\ket{0_{m}}\ket{1_{m}}+e_{2}\ket{1_{m}}\ket{0_{m}})_{a'b'} ,
\end{align}
where $c_{1}=c_{2}=e_{1}=e_{2}=1$. For generality we keep the coefficients named $c_{1},c_{2}$ and $e_{1},e_{2}$ since it will be important to consider $e_{1,2},c_{1,2} \neq 1$ in order to describe later swapping stages. We imagine combining modes $b$ and $a'$ on a balanced beamsplitter. Using the approximations \eqref{eq:approx1}, \eqref{eq:approx2}, we have the following transformations up to constants of $1/N_{\mu_{m}}^{\pm}$ and $1/N_{\tilde{\mu}_{m}}^{\pm}$ on the right-hand side 
\begin{align}
\ket{1_{m}}_{b}\ket{1_{m}}_{a'} \to & (|\sqrt{2}\mu_{m}\rangle_{b}\text{+}|\text{-}\sqrt{2}\mu_{m}\rangle_{b})\ket{0}_{a'} \nonumber \\
& + (|\sqrt{2}\mu_{m}\rangle_{a'}\text{+}|\text{-}\sqrt{2}\mu_{m}\rangle_{a'})\ket{0}_{b} \\
\ket{0_{m}}_{b}\ket{0_{m}}_{a'} \to & (|\sqrt{2}\tilde{\mu}_{m}\rangle_{b}\text{+}|\text{-}\sqrt{2}\tilde{\mu}_{m}\rangle_{b})\ket{0}_{a'} \nonumber \\
& - (|\sqrt{2}\tilde{\mu}_{m}\rangle_{a'}\text{+}|\text{-}\sqrt{2}\tilde{\mu}_{m}\rangle_{a'})\ket{0}_{b} \\
\ket{1_{m}}_{b}\ket{0_{m}}_{a'} \to & |(\tilde{\mu}_{m}\text{+}\mu_{m})/\sqrt{2}\rangle_{b}|(\text{-}\tilde{\mu}_{m}\text{+}\mu_{m})/\sqrt{2}\rangle_{a'} \nonumber \\
& - |(\text{-}\tilde{\mu}_{m}\text{+}\mu_{m})/\sqrt{2}\rangle_{b}|(\tilde{\mu}_{m}\text{+}\mu_{m})/\sqrt{2}\rangle_{a'} \nonumber \\
& + |(\tilde{\mu}_{m}\text{-}\mu_{m})/\sqrt{2}\rangle_{b}|(\text{-}\tilde{\mu}_{m}\text{-}\mu_{m})/\sqrt{2}\rangle_{a'} \nonumber \\
& - |(\text{-}\tilde{\mu}_{m}\text{-}\mu_{m})/\sqrt{2}\rangle_{b}|(\text{-}\tilde{\mu}_{m}\text{-}\mu_{m})/\sqrt{2}\rangle_{a'} \\
\ket{0_{m}}_{b}\ket{1_{m}}_{a'} \to & |(\tilde{\mu}_{m}\text{+}\mu_{m})/\sqrt{2}\rangle_{b}|(\text{-}\tilde{\mu}_{m}\text{+}\mu_{m})/\sqrt{2}\rangle_{a'} \nonumber \\
& + |(\text{-}\tilde{\mu}_{m}\text{+}\mu_{m})/\sqrt{2}\rangle_{b}|(\tilde{\mu}_{m}\text{+}\mu_{m})/\sqrt{2}\rangle_{a'} \nonumber \\
& - |(\tilde{\mu}_{m}\text{-}\mu_{m})/\sqrt{2}\rangle_{b}|(\text{-}\tilde{\mu}_{m}\text{-}\mu_{m})/\sqrt{2}\rangle_{a'} \nonumber \\
& - |(\text{-}\tilde{\mu}_{m}\text{-}\mu_{m})/\sqrt{2}\rangle_{b}|(\text{-}\tilde{\mu}_{m}\text{-}\mu_{m})/\sqrt{2}\rangle_{a'} .
\end{align} 
The squeezing operators $\hat{S}(2)_{b}\hat{S}(2)_{a'}$ should multiply the expressions on the right-hand side but we omit these for simplicity. Going to the wave function picture, assuming that $\mathrm{cos}(\sqrt{2}\mu_{m}p')\approx \mathrm{cos}(\sqrt{2}\tilde{\mu}_{m}p')$, $\mathrm{sin}(\sqrt{2}\mu_{m}p')\approx \mathrm{sin}(\sqrt{2}\tilde{\mu}_{m}p')$, and that $(1+e^{-2\mu_{m}^{2}})\approx(1-e^{-2\tilde{\mu}_{m}^{2}})$, we get the transformations 
\begin{align}
\ket{1_{m}}_{b}\ket{1_{m}}_{a'} & \to -2\mathrm{cos}(\sqrt{2}\mu_{m}p')e^{-p'^{2}-x^{2}}  \\
\ket{0_{m}}_{b}\ket{0_{m}}_{a'} & \to 2\mathrm{cos}(\sqrt{2}\mu_{m}p')e^{-p'^{2}-x^{2}}  \\
\ket{1_{m}}_{b}\ket{0_{m}}_{a'} & \to -i(e^{-\sqrt{2}x(\mu_{m}-\tilde{\mu}_{m})}+e^{\sqrt{2}x(\mu_{m}-\tilde{\mu}_{m})}) \times \nonumber \\
& e^{-\frac{1}{2}(\mu_{m}-\tilde{\mu}_{m})^{2}}\mathrm{sin}(\sqrt{2}\mu_{m}p')e^{-p'^{2}-x^{2}} \\
\ket{0_{m}}_{b}\ket{1_{m}}_{a'} & \to i(e^{-\sqrt{2}x(\mu_{m}-\tilde{\mu}_{m})}+e^{\sqrt{2}x(\mu_{m}-\tilde{\mu}_{m})}) \times \nonumber \\
& e^{-\frac{1}{2}(\mu_{m}-\tilde{\mu}_{m})^{2}}\mathrm{sin}(\sqrt{2}\mu_{m}p')e^{-p'^{2}-x^{2}} .
\end{align}
Here $p'$ is the momentum variable of mode $a'$ and $x$ is the position variable of mode $b$. We now perform the $\hat{X}$ measurement on mode $b$ and the $\hat{P}$ measurement on mode $a'$ and assume that we get outcomes $p'_{0}$ and $x_{0}$. The unnormalized state after the swapping is
\begin{align}
& e_{1}c_{2}A'\ket{0_{m}}_{a}\ket{0_{m}}_{b'}-e_{2}c_{1}A'\ket{1_{m}}_{a}\ket{1_{m}}_{b'} \nonumber \\
& +c_{1}c_{2}C\ket{1_{m}}_{a}\ket{0_{m}}_{b'}+e_{1}e_{2}C^{*}\ket{0_{m}}_{a}\ket{1_{m}}_{b'} ,
\end{align}
with
\begin{align}
A' & = 2\mathrm{cos}(\sqrt{2}\mu_{m}p'_{0}) \\
C' & = i(e^{-\sqrt{2}x_{0}(\mu_{m}-\tilde{\mu}_{m})}+e^{\sqrt{2}x_{0}(\mu_{m}-\tilde{\mu}_{m})}) \times \nonumber \\
& e^{-\frac{1}{2}(\mu_{m}-\tilde{\mu}_{m})^{2}}\mathrm{sin}(\sqrt{2}\mu_{m}p'_{0}) . 
\end{align}
Whenever $e_{i}=c_{i}^{*}$ this state contains one ebit of entanglement and can be written as:
\begin{align}
\label{eq:appswap1}
& A\ket{0_{m}}_{a}\ket{0_{m}}_{b'}-A^{*}\ket{1_{m}}_{a}\ket{1_{m}}_{b'} + \nonumber \\
& C\ket{1_{m}}_{a}\ket{0_{m}}_{b'}+C^{*}\ket{0_{m}}_{a}\ket{1_{m}}_{b'} ,
\end{align}
with
\begin{align}
\label{eq:coef1}
A & = c_{1}^{*}c_{2}2\mathrm{cos}(\sqrt{2}\mu_{m}p'_{0})\\
C & = c_{1}c_{2} i(e^{-\sqrt{2}x_{0}(\mu_{m}-\tilde{\mu}_{m})}+e^{\sqrt{2}x_{0}(\mu_{m}-\tilde{\mu}_{m})}) \times \nonumber \\
& e^{-\frac{1}{2}(\mu_{m}-\tilde{\mu}_{m})^{2}}\mathrm{sin}(\sqrt{2}\mu_{m}p'_{0}).
\end{align}
Swapping two states of the form in Eq.\eqref{eq:appswap1} with coefficients $A_{1},C_{1}$ and $A_{2},C_{2}$ respectively, it can be shown within the same approximations leading to Eq. \eqref{eq:appswap1} that the swapped state will also be of the form: 
\begin{align}
& A\ket{0_{m}}\ket{0_{m}}-A^{*}\ket{1_{m}}\ket{1_{m}} + \nonumber \\
& C\ket{1_{m}}\ket{0_{m}}+C^{*}\ket{0_{m}}\ket{1_{m}} ,
\end{align}
with coefficients
\begin{eqnarray}
A&=&\left[ \mathbb{A}\mathrm{cos}(\sqrt{2}\mu_{m}p'_{0})+\mathbb{B}\mathrm{sin}(\sqrt{2}\mu_{m}p'_{0}) \right] \nonumber \\
C&=&\left[ \mathbb{C}\mathrm{cos}(\sqrt{2}\mu_{m}p'_{0})+\mathbb{D}\mathrm{sin}(\sqrt{2}\mu_{m}p'_{0}) \right]
\end{eqnarray}
with $p'_{0}$ again being the outcome of the $\hat{P}$ measurement.
The coefficients $\mathbb{A},\mathbb{B},\mathbb{C}$ and $\mathbb{D}$ depend on the measurement outcome of the $\hat{X}$ measurement in the relevant swap and in the previous swap levels as well as the $\hat{P}$ measurement in the previous swap levels (see Eq \eqref{eq:coef1}). In the simulation of the repeater we replace $\sqrt{2}\mu_{m}p'_{0} \to \theta_{p} $ and optimize the fidelity between the swapped state and the target state with respect to $\theta_{p}$. Note that regardless of this we always calculate the fidelity with a pure state containing one ebit of entanglement. 

\section{Parameters of fidelity fits} \label{app:supma}

In this appendix we list the parameters of the fidelity fits shown in \tabref{tab:table2} of the article. The matrices below contain the constants $\tilde{a}_{n,m}$-$\tilde{h}_{n,m}$. The first entry in a matrix is for $n=0,m=1$ and so fourth.  
\begin{align*}
\mathbf{a}&=
\begin{pmatrix}
0 & 0 & 0 \\
0 & 0 & 0 \\
-2.19 & -5.39 & -6.81 \\
-9.75 & -14.6 & -20.1 \\
-15.6 & -26.1 & -39.9
\end{pmatrix}   \\
\mathbf{b}&=
\begin{pmatrix}
0.90 & 0.91 & 0.95 \\
1.40 & 1.53 & 1.65 \\
2.25 & 3.08 & 3.40 \\
3.69 & 4.92 & 5.83 \\
4.26 & 6.46 & 8.54
\end{pmatrix}
\end{align*}
\begin{align*}
\mathbf{c}&= 
\begin{pmatrix}
0.0063 & 1.0 & 4.7 \\
0.223 & 1.50 & 5.08 \\
0.460 & 2.59 & 6.26 \\
1.56 & 3.73 & 8.77 \\
2.02 & 6.68 & 16.1
\end{pmatrix}  \cdot10^{-3}  \\
\mathbf{d}&=
\begin{pmatrix}
15.0 & 24.2 & 92.0 \\
13.1 & 23.1 & 93.8 \\
12.3 & 21.6 & 94.5 \\
10.2 & 21.0 & 92.6 \\
9.47 & 19.0 & 83.8
\end{pmatrix}  \\
\mathbf{e}&=
\begin{pmatrix}
- & - & - \\
-466 & -8.90 \cdot 10^{-7} & 0.993\\
0.969 & -2.71 \cdot 10^{-6} & 0.979 \\
1.32 & -1.93 \cdot 10^{-5} & 0.954 \\
0.824 & -1.74 \cdot 10^{-4} & 0.905
\end{pmatrix} \\
\mathbf{f}&=
\begin{pmatrix}
- & - & - \\
0.351 & 5.39 & 0.411 \cdot 10^{-3}\\
0.324 & 5.32 & 0.792\cdot 10^{-3} \\
-0.592 & 4.58 & 3.81\cdot 10^{-3} \\
-0.411 & 3.60& 5.30\cdot 10^{-3}
\end{pmatrix}  \\
\end{align*} 
\begin{align*}
\mathbf{g}&=
\begin{pmatrix}
- & - & - \\
468 & 0.985 & -\\
-0.969 & 0.951 & - \\
-0.636 & 0.893 & - \\
-0.260 & 0.799 & -
\end{pmatrix}  \\
\mathbf{h}&=
\begin{pmatrix}
- & - & - \\
0.350 & -2.80 \cdot 10^{-3}& -\\
0.324 &  1.98\cdot 10^{-3}& - \\
-2.45 & 3.78\cdot 10^{-3} & - \\
-4.24 & 7.20\cdot 10^{-3} & -
\end{pmatrix} 
\end{align*} 
Matrices $\mathbf{e}-\mathbf{h}$ show that the state's swap performance increases for large values of $m$ and matrices $\mathbf{a}-\mathbf{d}$ show that the fidelity drops as a function of $r$ and $\vec{\Delta}$. The fact that $\tilde{a}_{n,m} \le 0$ reflects that the fits are made for $P_{connect} \ll1$.

The numerical vectors of the constants $\tilde{i}_{n}$-$\tilde{k}_{n}$ and $\tilde{l}_{m}$ are
\begin{eqnarray}
 \mathbf{i}&=& \left(
 \begin{array}{ccccc} 1& 0.938& 0.811& 0.618&0.413\end{array} \right) \nonumber  \\
  \mathbf{j}&=& \left(
 \begin{array}{ccccc} 0& -0.460  & -1.61& -3.14 & -3.64\end{array} \right) \cdot 10^{-3} \qquad \nonumber \\
  \mathbf{k}&=& \left(
 \begin{array}{ccccc} 0& 0.0323& 0.104& 0.207&0.275\end{array} \right)  \nonumber \\
  \mathbf{l}&=& \left(
 \begin{array}{ccc} 0.0112& 0.0222& 0.00542\end{array} \right). \nonumber
\end{eqnarray}
Vectors $\mathbf{i}-\mathbf{k}$ show that the output fidelity drops as a function of $n$ and vector $\mathbf{l}$ shows that as $m$ increases the states gets more robust to the swapping procedure. 

\end{document}